\documentclass[]{jfm}
\usepackage{graphicx}
\usepackage{epstopdf, epsfig}
\usepackage{amsmath, amssymb}
\usepackage{bm}
\usepackage{here}
\usepackage{graphicx,siunitx}
\usepackage{graphicx}
\usepackage{listings,color,courier,natbib}
\usepackage{pdflscape, csquotes}
\graphicspath{{./figures/}}
\usepackage[abs]{overpic}

\usepackage[T1]{fontenc}
\usepackage{adjustbox}
\usepackage{rotating}

\shorttitle{Super-resolution reconstruction with machine learning}
\shortauthor{K. Fukami, K. Fukagata and K. Taira}

\title{Super-resolution reconstruction of turbulent flows with machine learning}

\author{Kai Fukami\aff{1,2,3}
  \corresp{\email{kai.fukami@keio.jp}},
  Koji Fukagata\aff{1}
 \and Kunihiko Taira\aff{2,3}}

\affiliation{\aff{1}Department of Mechanical Engineering, Keio University, Yokohama, 223-8522, Japan
\aff{2}Department of Mechanical Engineering, Florida State University, Tallahassee, FL 32310, USA
\aff{3}Department of Mechanical and Aerospace Engineering, University of California, Los Angeles, CA 90095, USA
}

\begin{document}

\maketitle

\begin{abstract}
We use machine learning to perform super-resolution analysis of grossly under-resolved turbulent flow field data to reconstruct the high-resolution flow field.  Two machine-learning models are developed; namely the convolutional neural network (CNN) and the hybrid Downsampled Skip-Connection Multi-Scale (DSC/MS) models. 
These machine-learning models are applied to two-dimensional cylinder wake as a preliminary test and show remarkable ability to reconstruct laminar flow from low-resolution flow field data.  We further assess the performance of these models for two-dimensional homogeneous turbulence.  The CNN and DSC/MS models are found to reconstruct turbulent flows from extremely coarse flow field images with remarkable accuracy.  For the turbulent flow problem, the machine-leaning based super-resolution analysis can greatly enhance the spatial resolution with as little as 50 training snapshot data, holding great potential to reveal subgrid-scale physics of complex turbulent flows.  With the growing availability of flow field data from high-fidelity simulations and experiments, the present approach motivates the development of effective super-resolution models for a variety of fluid flows. 
\end{abstract}


\section{Introduction}
The quest for high-resolution flow data has been one of the major pursuits in both experimental and computational fluid dynamics. The miniaturization of hot wires and advancement in particle image velocimetry technology have revealed intricate details of turbulent flow structures. On the computational side, the increasing spatial grid resolution that a computer can handle has enabled high-fidelity simulations to uncover the richness of turbulence.  With the explosion in the size of collected fluid flow data, we expect that the information contained therein can complement these experimental and computational endeavors by taking advantage of the powerful machine learning techniques. In this study, we capitalize on machine learning to reconstruct unsteady laminar and turbulent flows from spatially low-resolution data.

In recent years, machine learning (ML) has emerged as a promising technique to develop turbulence models for various applications \citep{Ling2016,Kutz2016}. \citet{DIX2019} developed accurate closure models for Reynolds-Averaged Navier--Stokes (RANS) using multi-layer percepton type neural network. \citet{Ling2016} also proposed RANS modeling using a tensor-basis neural network with Galilean invariance embedded. They tested the ML model for duct and wavy-wall flows. 
\citet{SM2018} used the ML architecture for reduced-order modeling of turbulent systems and showed its advantage against the proper orthogonal decomposition based model.
\citet{MS2017} also utilized the blind deconvolution method in large-eddy simulation using multi-layer perceptron to estimate the eddy viscosity. Their results show statistical agreement with the reference data. 
Moreover, a machine-learned turbulence generator has been developed by \citet{FKF2018}.
This machine-learned turbulence generator reduces the computation time by approximately 150 times against traditional direct numerical simulation (DNS) based turbulence generator while maintaining the turbulent statistics. 
Machine learning has also been utilized for feedback control by \citet{KTS2018}. In their work, reinforcement learning with a deep Q-network was used to perform closed-loop cylinder wake control, achieving a $34\%$ drag reduction.

In addition to the modeling effort, there is a critical need for data reconstruction in general that can benefit from machine learning techniques.
The fluid dynamics field is no exception. \citet{LMB2018} inferred the rotation rate and temperature of turbulent flow by using spectral nudging and showed reasonable agreement with the reference DNS at low Reynolds number.
In the field of computer science, reconstruction of high-resolution (HR) images from low-resolution (LR) images has been an active area of research. 
Noteworthy here is the development of facial reconstruction technology from coarse images. 
Such advancement has broad implications ranging from data compression, communications, and security. The reconstruction of images based on their low-resolution data is known as {\it super-resolution analysis}. 
Bicubic interpolation is one of the traditional super-resolution methodologies based on the filter operation which has low-pass characteristics \citep{Keys1981}. Although the implementation of such algorithm is easy, it is not suitable for high-frequency reconstruction. 
To address this issue, there has been emerging efforts on adopting ML to perform super-resolution analysis, achieving remarkable results in the image tasks.
\citet{DLHT2015} proposed the super-resolution convolutional neural network (SRCNN) as the pioneer of machine-learning based super-resolution analysis.
In this study, we examine the use of machine-learned super-resolution analysis of low-resolution complex fluid flow images.

The objective of the present work is to demonstrate that machine-learned techniques can reconstruct high-resolution flow fields from low-resolution images.
Convolutional neural network (CNN) model and the hybrid Downsampled Skip-Connection Multi-Scale (DSC/MS) model are used in the machine-learned super-resolution analysis, as presented in \S 2.
We use the two-dimensional cylinder wake and two-dimensional homogeneous turbulence as test cases in \S 3.
At last, we discuss some of the key findings to accurately reconstruct the multi-scale turbulent flows from super-low-resolution images.
We offer some concluding remarks in \S 4.

\section{Methodology}

Given the input data set $\bm x$ $\in {\mathbb{R}^r}$ and the desired output data set $\bm y$ $\in {\mathbb{R}^n}$, we aim to find the optimal weight $\bm w$ in a machine-learned model $\cal F$ that acts as a nonlinear regression function such that ${\cal F}({\bm x}; {\bm w}) \approx {\bm y}$. In the present case, ${\bm x}$ and ${\cal F}({\bm x}; {\bm w})$ represent the low-resolution and reconstructed high-resolution data, respectively. 
The weight $\bm w$ is optimized such that the $L_2$ norm between the desired high-resolution output $\bm y$ and the ML model output ${\cal F}({\bm x}; {\bm w})$ is minimized, i.e.,
\begin{equation}
    {\bm w} =  \text{argmin}_{\bm w} \|{\bm y}-{\cal F}({\bm x}; {\bm w})\|_2^2 .
    \label{eq:weights}
\end{equation}
Once $\bm w$ is determined from the training data, the ML model $\mathcal{F}$ is ready for use.

\begin{figure}
	\begin{center}
		\includegraphics[width=0.95\textwidth]{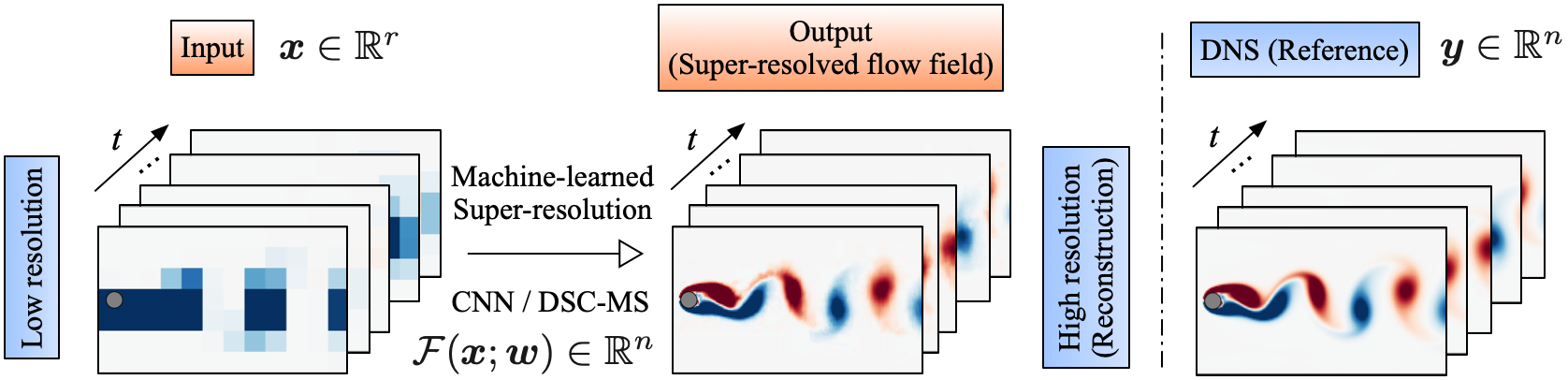}
		\caption{An overview of machine-learned super-resolution analysis for cylinder flow.}
		\label{fig1}
	\end{center}
\end{figure}

The overall procedure of machine-learned super-resolution analysis is presented in figure \ref{fig1}. As an illustration, we show the application of the ML model on two-dimensional laminar cylinder flow as discussed later in details.
The reference data sets are obtained by direct numerical simulation (DNS). The low-resolution data is fed to the machine-learned model and then attempts to reconstruct the flow fields.

\begin{figure}
	\begin{center}
		\includegraphics[width=0.8\textwidth]{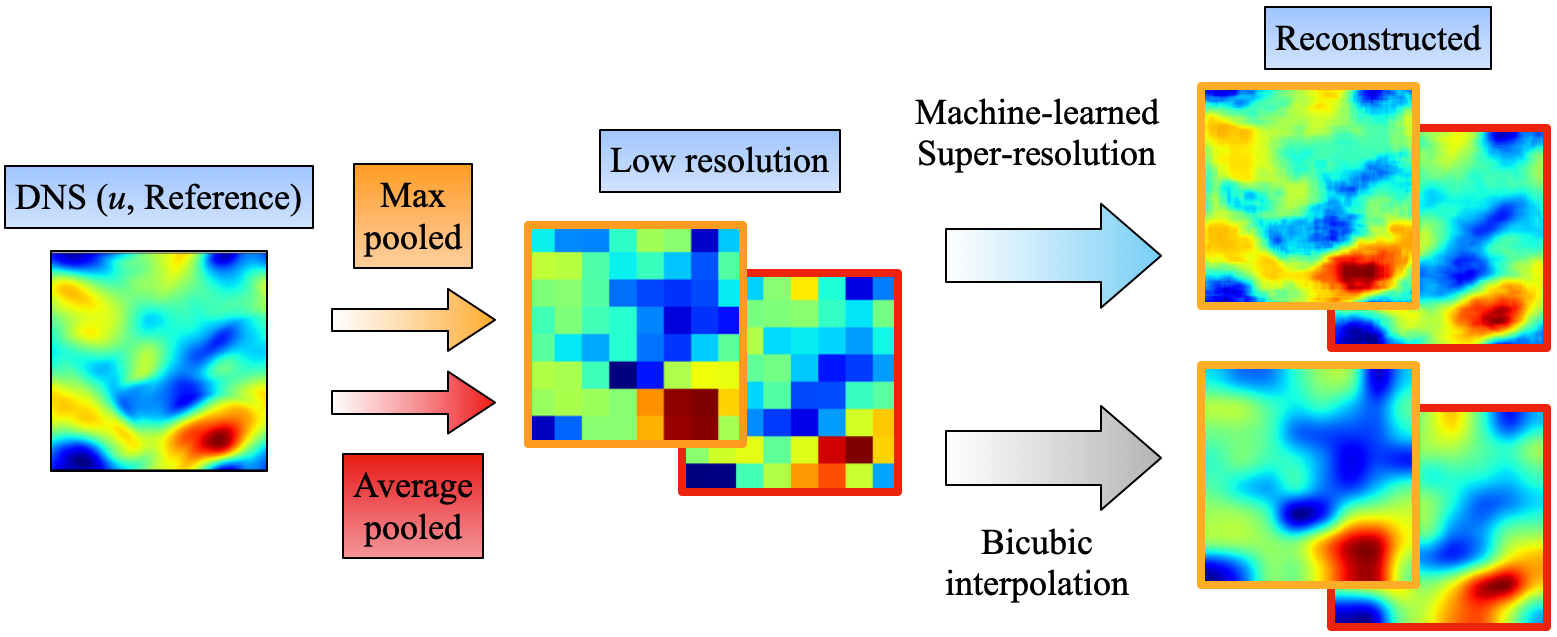}
		\caption{
		Max/average pooling and super-resolution reconstruction of the turbulent velocity field.}
		\label{fig3}
	\end{center}
\end{figure}

For the purpose of preparing the input data set, we choose average and max pooling to downsample the original DNS flow field, as illustrated in figure \ref{fig3}.  These pooling operations are defined as 
\begin{equation}
 	q_{ij}^\text{LR} = \bigg[ \frac{1}{M^2} \sum_{p,s \in {P}_{i,j}} \left( q_{ps}^\text{HR} \right)^{P} \bigg]^{1/{P}},
\end{equation}
where $P = 1$ and $\infty$ provide average and max pooling, respectively, over a square pooling window of $M \times M$ pixels.  This enables the original image of size $L_{\alpha} \times L_{\beta}$ pixels to be reduced to $(L_{\alpha}/M) \times (L_{\beta}/M)$ pixels.  In terms of the variables sizes in equation \ref{eq:weights}, we have $n=L_\alpha \times L_\beta \times K$ and $r=(L_{\alpha}/M) \times (L_{\beta}/M) \times K$, where we take $L_{\alpha} = N_x$ and $L_{\beta} = N_y$.
In what follows, we choose $M$ to be $8$ (medium-resolution (MR)), $16$ (low-resolution (LR)) and $32$ (super-low-resolution (SLR)) images, respectively.  The above two pooling procedures are considered in this study due to their widespread use in image processing.  The average pooling is a simple arithmetic averaging operation that is encountered in common downsampling. On the other hand, max pooling is widely adopted in image processing to retain the range of color and brightness. The difference in the downsampled image of the turbulent flow can be seen in figure \ref{fig3}. The regions with large magnitudes of $u$ velocity component are well captured by max pooling, while the flow field appears smoothed with average pooling.

In the present work, two ML models are examined for developing $\cal F$.  As the first ML model, we consider a Convolutional Neural Network (CNN) model that is widely used for image recognition \citep{LBBH1998}, including super-resolution analysis \citep{DLHT2015, RIM2016}. 
Recently, the CNN has shown its strength, especially for handling big fluid flow data \citep{ZSM2018, HDE2019, SP2019}.

\begin{figure}
	\begin{center}
		\includegraphics[width=1.0\textwidth]{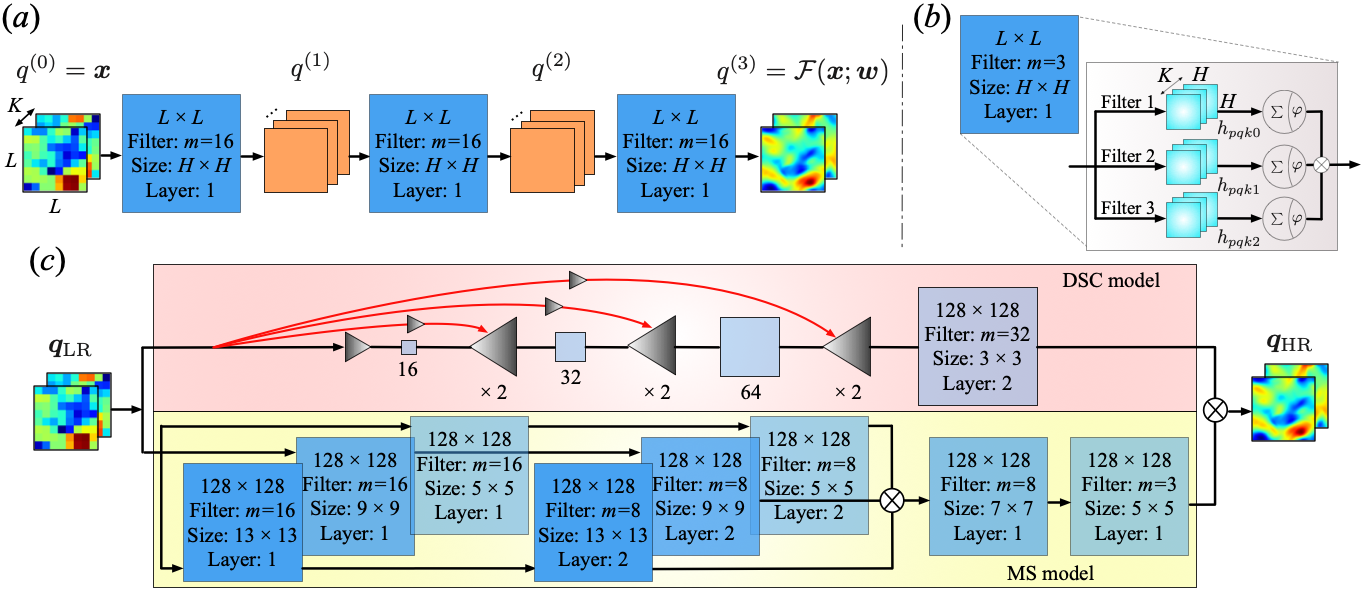}
		\vspace{-3mm}
		\caption{$(a)$ Schematic of the convolutional neural network (CNN) with two-dimensional turbulent flow. $(b)$ Inner working of each CNN, shown for a 3-filter setup.
		$(c)$ Schematic of the hybrid Downsampled Skip-Connection Multi-Scale (DSC/MS) model.}
		\label{fig2}
	\end{center}
\end{figure}

CNN processes the input data in an iterative manner from variable $q^{(l-1)}$ to $q^{(l)}$.  Starting with the input $q^{(0)} = \bm{x}$, we have  
\begin{equation}
q^{(l)}_{ijm} = {\varphi}\biggl(\sum^{K-1}_{k=0}\sum^{L-1}_{p=0}\sum^{L-1}_{s=0}q_{i+p,j+{s},k}^{(l-1)}h_{p{s}km}\biggr),
\end{equation}
where $q^{(l_\text{max})} = {\mathcal F}({\bm x}; {\bm w})$.  In the above formula,
$q^{(l)}$ and $q^{(l-1)}$ are the input and output variables, respectively, for layer $l$.  This procedure with two-dimensional turbulent flow is illustrated in figure \ref{fig2}$(a)$ for a three-layer ($l_\text{max} = 3$) CNN model with an example layer expanded in figure \ref{fig2}$(b)$.  In the diagram, $H$ is the length of the filter $h$, $\varphi$ represents the activation function,  $L$ (= $L_\alpha$ = $L_\beta$, for the example of two-dimensional turbulence) is the number of pixels in each direction, and $K$ denotes the number of images constituting the data (e.g., for color images, $K=3$ for the RGB (red, green and blue) code).   
We have used a filter which incorporates the periodic boundary condition into the padding operation. However, there were no significant differences in the results from the use of zero padding, which is commonly used for image processing.
 Moreover, we use the rectified linear unit (ReLU) $f(z) = {\rm max}(0,z)$ as the activation function $\varphi$.
The use of ReLU is known to be an effective tool for stabilizing the weight update in machine learning process \citep{NH2010}.
The weights of the filters ${\bm w}$ are determined using the Adaptive Moment Estimation (Adam) algorithm \citep{KB2014}.

To perform the super-resolution analysis of fluid flows, we select the velocity vector $\bm{u} = \{u,v\}$ or the vorticity field $\omega$ as the input, making $K=2$ or $1$, respectively. We provide the velocity and vorticity fields separately to analyze their respective influence on the accuracy of the super-resolution analysis.  There are some important differences for the use of these two inputs.  First is the difference between a vector and a scalar input to the model.  Another is the difference in the spectral components over the high-wavenumbers.  The vorticity field is amplified by the wavenumber in comparison to the velocity field, which serves as a nice test for the machine learning process.

The above CNN model can be further improved to perform super-resolution reconstruction of a coarse fluid flow data.  In particular, we consider a second approach that is a hybrid of two techniques to capture both large and small-scale structures, which is ideal for turbulence.  First, we extend the CNN model by introducing compression and skipped connections, as shown in the red box of figure \ref{fig2}$(c)$.  In super-resolution analysis, data compression (triangular operations) increases the robustness against translation and rotation of the data elements \citep{LNCCK2010}.  The use of skipped connections (red arrows) enhances the CNN prediction by removing issues related to the convergence of weights \citep{HZRS2015} which is known to be a problem with deep CNNs.
We also introduce the multi-scale model by \citet{DQHG2018} that captures the small-scale structures in the data.  This multi-scale model is shown in the yellow box of figure \ref{fig2}$(c)$ and is comprised of a number of CNN filters with different lengths to span a range of scales.  The extended super-resolution approach combines the DSC and MS models, and is referred to as the hybrid Downsampled Skip-Connection/Multi-Scale (DSC/MS) model\footnote{The source code for the hybrid DSC/MS model presented in this study will be made available online at the time of publication of this article.}.  While the discussion is kept brief here, the full description of the methodology is presented in figure \ref{fig2}$(c)$.

For both of these ML models, we apply the early stopping criterion with 20 learning iterations to avoid an overfitting \citep{Prechelt1998}.  In what follows, we compare the performance of super-resolution analysis using the simple bicubic interpolation of coarse data, CNN reconstruction, and hybrid DSC/MS model based reconstruction on the laminar cylinder wake and the canonical decaying homogeneous two-dimensional turbulent flow fields.  
In what follows, we present the details on the ML reconstruction approach and demonstrate its validity for the reconstruction of complex fluid flows.

\section{Results}
\subsection{Example 1: Two-dimensional cylinder wake}

As a preliminary test of machine-learned super-resolution analysis, we consider the two-dimensional cylinder wake flow at $Re_D = 100$ \citep{TC2007, CT2008}.
The governing equations are the incompressible Navier-Stokes equations,
\begin{eqnarray}
\nabla \cdot {\bm u}= 0, \\
\dfrac{\partial {\bm u}}{\partial t}=-{\bm u} \cdot \nabla {\bm u}-\nabla p+\dfrac{1}{Re_D}\nabla^2\bm u,
\end{eqnarray}
where $\bm u$, $p$ and $Re_D$ are the non-dimensionalized velocity vector, pressure and Reynolds number, respectively.
The size of computational domain, the number of grid points and the range of time-steps are ($x/D$, $y/D$) = $[-0.7, 15] \times [-5, 5]$, ($N_x$, $N_y$) = (192, 112) and $\Delta t=2.50\times 10^{-3}$.
In this particular demonstration, the vorticity field $\omega$ is used as the input and output attributes. 
Also we use the max pooling operation of low-resolution for obtaining the coarse data.
The reconstructed vorticity color contour is shown in figure \ref{fig1}. This machine learning model is made using 1000 snapshots corresponding to 8 cycles of a wake flow.

\begin{figure}
	\begin{center}
		\includegraphics[width=0.90\textwidth]{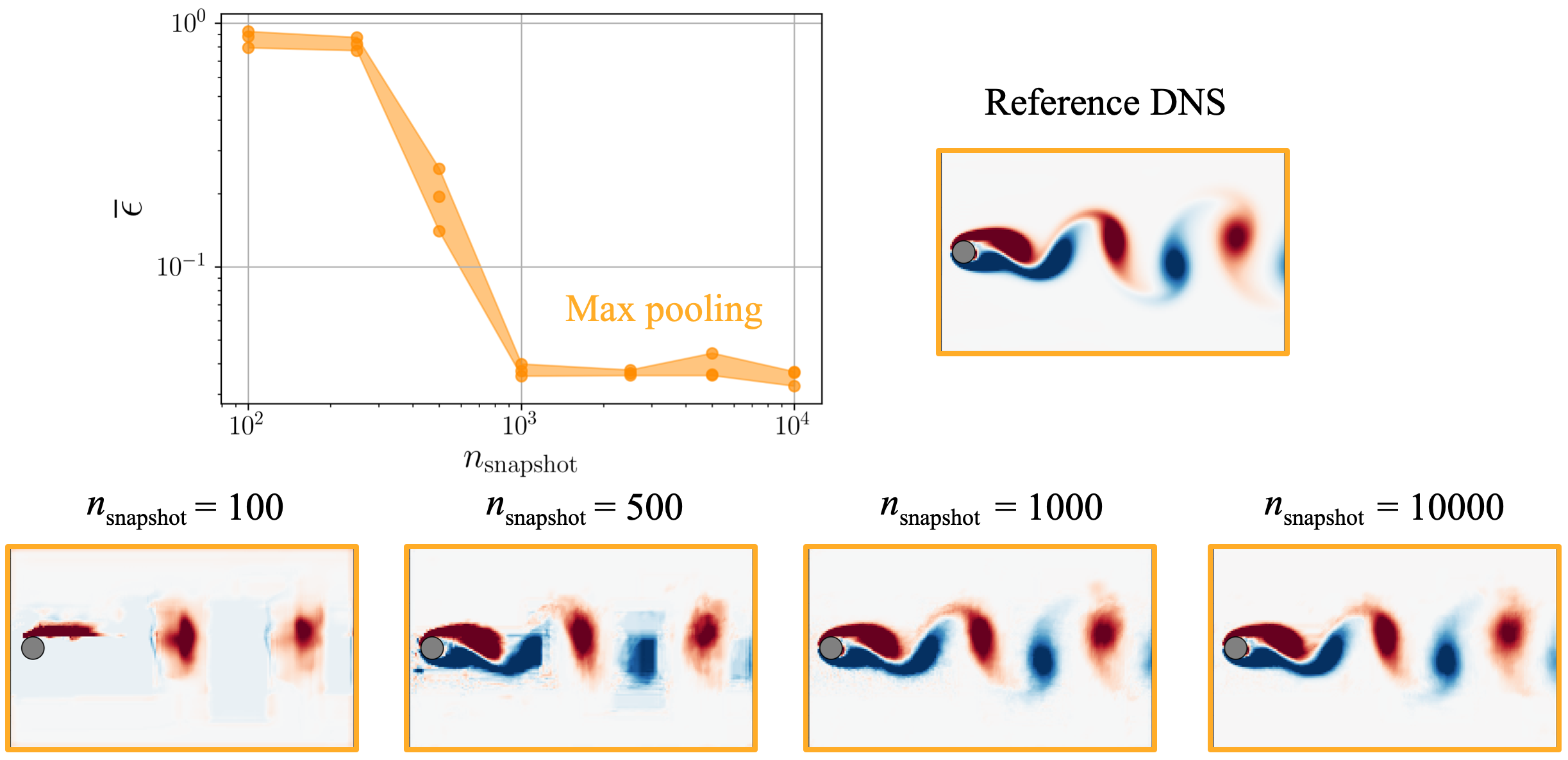}
        \caption{The dependence of the error $\overline{\epsilon}$ on the number of snapshots $n_{\rm snapshot}$ for the laminar cylinder wake at $Re_D=100$. Shown on the right are the reconstructed vorticity flow from low-resolution input data sets. }
		\label{fig1_1}
	\end{center}
\end{figure}

\begin{figure}
	\begin{center}
		\includegraphics[width=0.50\textwidth]{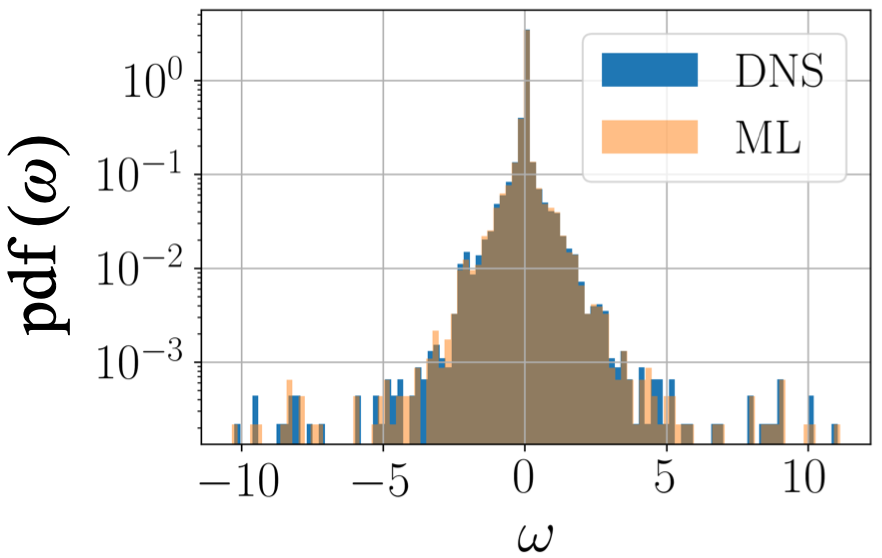}
        \caption{Probability density function of vorticity field pdf$(\omega)$ of laminar cylinder wake.}
		\label{fig1_2}
	\end{center}
\end{figure}

These show good agreement with the reference DNS data. The dependence of the number of snapshots $n_{\rm snapshot}$ in terms of $L_2$ error norm is assessed as shown in figure \ref{fig1_1}. These coarse snapshots are selected at even time intervals from the training data sets. It can be seen that $n_{\rm snapshot}=1000$ is sufficient to recover the flow field from the coarse data.
The probability density function of the vorticity field pdf$(\omega)$ is shown in figure \ref{fig1_2} exhibiting great agreement with the reference vorticity field pdf$(\omega$). The flow field is recovered well by using the machine-learned super-resolution analysis.
Note that, we assess using the test data which is not included in the leaning process.  From these observations, we confirm the effectiveness of the machine-learned super-resolution analysis for laminar flow.

\subsection{Example 2: Two-dimensional decaying isotropic turbulence}

To demonstrate the capability of the machine-learned super resolution reconstruction, we consider the two-dimensional homogeneous decaying turbulent flow simulated by a bi-periodic Fourier spectral incompressible direct numerical simulation (DNS) solver \citep{TNB2016}.
The reference flow field is obtained by numerically solving the two-dimensional vorticity transport equation
\begin{equation}
\frac{\partial \omega}{\partial t}+{\bm u}\cdot\nabla\omega=\frac{1}{Re_0}\nabla^2 \omega,
\label{eq_1}
\end{equation} 
where ${\bm u}=(u,v)$ and $\omega$ are the velocity and vorticity variables, respectively.  The size of the computational domain and the numbers of grid points are $L_x=L_y=1$ and $N_x=N_y=128$, respectively. The Reynolds number is defined as $Re \equiv u^*l^*/\nu$, where $u^*$ is the characteristic velocity given by the square root of the spatially averaged initial kinetic energy, $l^*$ is the initial integral length, and $\nu$ is the kinematic viscosity.  In our study, two-dimensional turbulent flows are initialized randomly such that the initial Reynolds number $Re_0 = u^*(t_0)l^*(t_0)/\nu=74.6$ (for training/validation data) and 87.7 (for test data).
The training/validation data sets spanning over $0.195 \le t \le 2.145$ are used for constructing the machine learning model and the test data sets which is not included in the learning process are used for the assessment of the model. The present flow field data exhibits both direct and inverse cascades over time.
To examine the turbulent flow data, we generate a collection of snapshots using DNS, and 70\% of which are used for training, while the remaining 30\% are used for validation. As the model input and output attributes, we use the velocity vector ${\bm u}$ or the vorticity $\omega$ to assess their influence on the accuracy of the machine-learned model.  We mention in passing that the difference in initial Reynolds number changes the overall statistics of the flow in a modest manner.  We test our machine-learned model to see their  ability to operate in a regime similar to the training data.  The validation process randomly selects some snapshots, which automatically accounts for the changing Reynolds number within the provided data.

To demonstrate ML-based super-resolution reconstruction of coarse turbulent flow data, we consider the applications of the CNN and hybrid DSC/MS models.
For comparison, we also utilize a simple bicubic interpolation, which is a traditional super-resolution method in the field of image tasks \citep{Keys1981}. Here, we take coarse velocity $\bm{u}$ and vorticity $\omega$ field data from the two-dimensional homogeneous turbulent flow and reconstruct the low-resolution flow on a higher resolution grid using the ML approaches of the CNN and hybrid DSC/MS models.
Once the coarse flow is reconstructed, we compare the super-resolution reconstruction with the original flow field from DNS and the reconstruction based on the bicubic interpolation.

\begin{figure}
\begin{center}
    \includegraphics[width=1.0\textwidth]{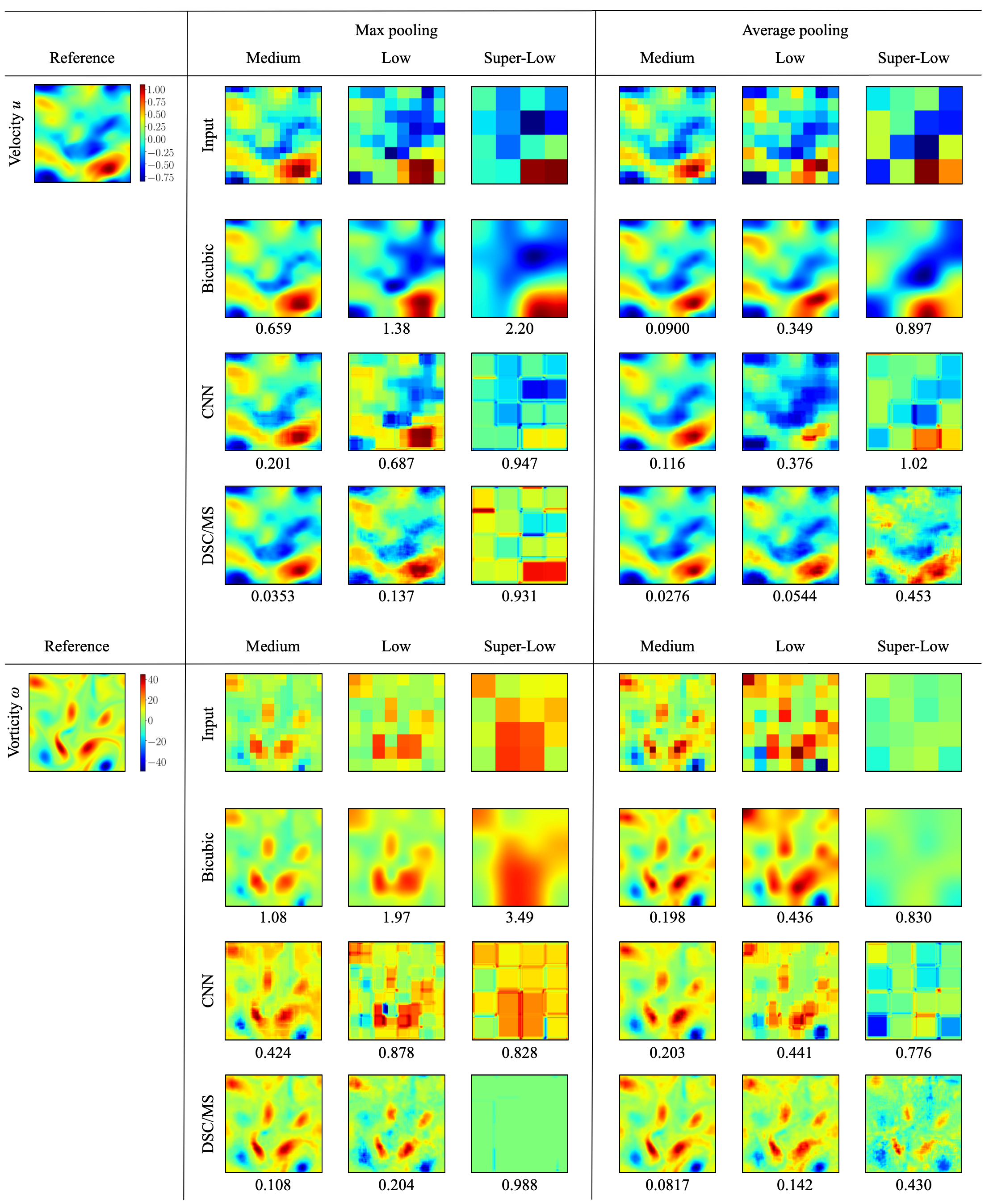}
	\caption{The color contours of velocity $u$ and vorticity $\omega$ in two-dimensional turbulent flow reconstructions with machine-learned super-resolution. Listed values indicate the $L_2$ error $\epsilon$.}
	\label{fig4}
	\end{center}
\end{figure}

Let us present a summary of results obtained by using $n_{\rm snapshot}=10\,000$ to highlight the capability of the super-resolution reconstruction of turbulent flow, in figure \ref{fig4}. 
Shown are the reconstructed $u$-component field using the velocity vector data ${\bm u}$ on the top half, and those using the scalar vorticity field $\omega$ on the bottom half of the figure. Both of the shown data sets are from the same time $t$ for a canonical turbulent flow field. The flows on the left and right halves correspond to the results using max and average pooled data sets, respectively, as inputs.  For each of the cases, the top rows represent the coarse input data to the ML models with varied resolutions. The following three rows present the super-resolution reconstructions by the bicubic, CNN, and hybrid DSC/MS models. 
The $L_2$ error norms, $\epsilon \equiv \|{\bm x}^\text{HR}-\mathcal{F}({\bm x})\|_2/\|{\bm x}^\text{HR}\|_2$, are reported underneath the reconstructed flow fields.  Note that this error norm is a strict measure of difference and does not account for translational or rotational similarities.  

Let us first examine the reconstruction of the velocity field with the simple bicubic interpolation.  For the medium-resolution velocity field, we find that the interpolation routine exhibits qualitative agreement with the reference flow field, but with a sizeable error of $0.659$ due to the inability to reconstruct the fine-scale structures.  The application of the bicubic interpolation to the medium-resolution vorticity field yields a higher level of error of $1.08$, as the vorticity field contains even finer scale structures.  As the input data is further coarsened, we find that the bicubic interpolation oversmoothes the flow field.  In fact, such gross regularizations are especially evident for the super-low cases for both the velocity and vorticity data inputs.  While the bicubic interpolation has difficulty reconstructing the flow in general, we notice that it performs better for the average pooled data sets compared to the max pooled cases with approximately five-fold error reduction.

The reconstruction of the flow field from the max pooled input data can be improved with the use of the CNN-based super-resolution analysis.  We can observe both for the velocity and vorticity field cases, the error levels are reduced noticeably.  Although the results appear pixelized, the error levels are indeed lowered, as the error is minimized through our choice of the cost function (\ref{eq:weights}).  For the average pooled input data, the CNN models attain super-resolution performance comparable to the bicubic interpolation.  These trends are shared for cases where the velocity and vorticity fields are provided.  In order to enhance the results with machine learning, we can suspect that a multi-scale connection may be necessary within the ML architecture based on the  limitations.

\begin{figure}
	\begin{center}
		\includegraphics[width=1\textwidth]{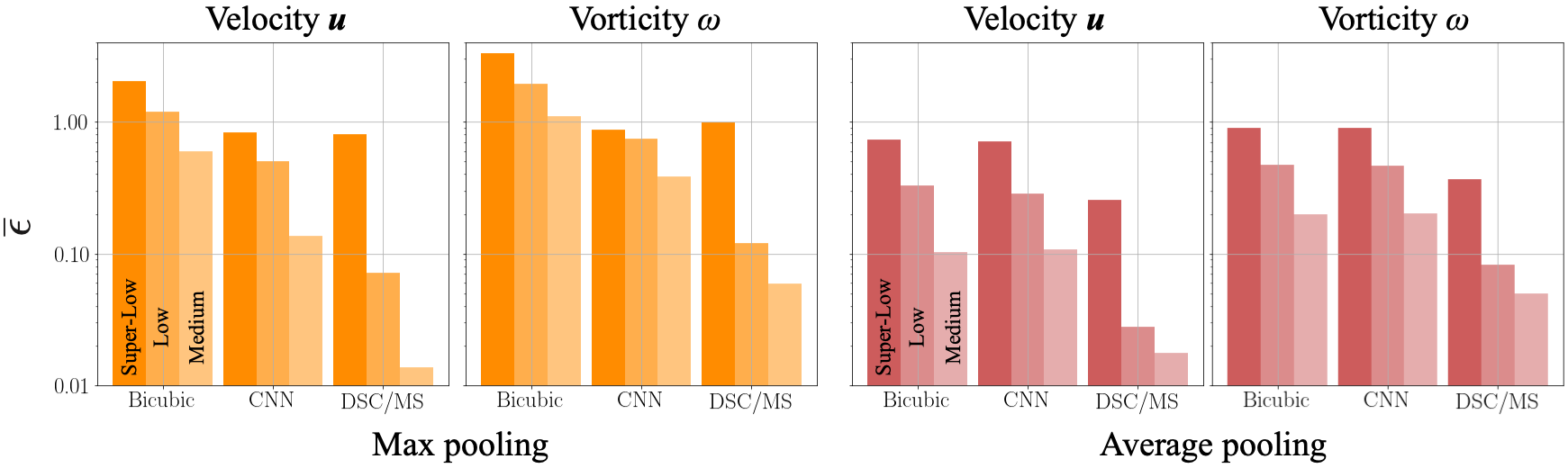}
		\caption{The ensemble-averaged super-resolution error $\overline{\epsilon}$ of the reconstructed flow using the max and average pooled velocity and vorticity data input.}
		\label{fig5}
	\end{center}
\end{figure}

To further enhance the reconstruction of fine-scale structures from the under-resolved input data, we consider the hybrid DSC/MS super-resolution model.  Compared to the reconstructed flows from the bicubic interpolation and CNN, the hybrid DSC/MS model in general recovers the turbulent flow with significant error reduction.  The only difficulty encountered by the hybrid DSC/MS model is for the super-low max pooled input data.  The application of max pooling to the training data appears to make the learning process difficult for super resolution.  The hybrid DSC/MS shows the noteworthy difference from the other two models in the reconstruction of the super-low average-pooled flow field.  Even with the given extremely low resolution of 4 $\times$ 4, the general distributions of the velocity and vorticity fields are recovered on the 128 $\times$ 128 grid.  Due to way the hybrid DSC/MS model processes multi-scale structures in the input data, its super-resolution process is able to reconstruct the flow in greater details.  The findings here suggest that the training data for ML models can be very coarse but should not come from max pooled data. This implies that use of coarse graphical images as input should be carefully considered, especially on how the low-resolution images were collected.
We note in passing that the present input data are not normalized enabling the models to be scale-invariant in outputting the reconstructed flow fields.

The results shown in figure \ref{fig4} were concerned with the super-resolution reconstruction of a single canonical turbulent flow field.  Next, let us evaluate the ensemble average of the $L_2$ error norm $\epsilon$ for the velocity and vorticity fields.  Presented in figure \ref{fig5} are the errors evaluated over 2000 test snapshots excluded from the learning process.  As already mentioned, we find that the na\"ive bicubic interpolation cannot accurately reconstruct the turbulent, especially with max pooled data.  The CNN-based super-resolution reconstruction performs better for the max pooled data set but does not reduce the error level for the average pooled input.  What is striking in the error comparison is the remarkable performance of super-resolution reconstruction by the hybrid DSC/MS model, which almost always outperforms the other models.  The performance of all models are lowered for the vorticity field due to its spectral content increasing over the high wavenumbers.  We note in passing that the super-low-resolution input is close to the limit of recovering the original flow field, which will be discussed further later. 

\begin{table}
\begin{center}
\def~{\hphantom{0}}
\begin{tabular}{cccccc}
     		   & Bicubic & CNN & DSC/MS \\
     Training [{\rm h}] & --- & $4.05 $ & $6.96 $\\
      & & $=18~{\rm s/ep.} \times 809~{\rm ep.}$ & $=96~{\rm s/ep.} \times 261~{\rm ep.}$ \\
     Reconstruction [{\rm s}] & $6.69\times10^{-2}$&$2.65\times10^{-3}$&$1.32\times10^{-2}$ \\
    \end{tabular}
    \\
  \caption{Computation time for training the machine-learned models and reconstructing 1 snapshot. As an example case, we use the low-resolution velocity vector model with average pooling and indicates 10\,000 snapshots. Epochs (ep.) indicates the number of iterations used in the learning process.}
  \label{tab1}
\end{center}
\end{table}

The computation times for training the machine-learned models and reconstructing a snapshot are summarized in table \ref{tab1}. The low-resolution velocity vector models with the average pooled data are chosen with 10\,000 snapshots for training in this example. To derive the CNN and hybrid DSC/MS models, approximately 4 and 7 hours, respectively, are needed using the NVIDIA Tesla K40 graphics processing unit (GPU). The reconstruction time for a single snapshot with the same GPU are also shown in table \ref{tab1}. Both machine-learned models are able to reconstruct a flow field faster than the bicubic interpolation. The above reported times are for an input data of $128\times128$ pixels in the current paper.  For a data size of $256\times256$ pixels, the computation time for a single learning iteration (referred to as an epoch) increases by 6 times.  While this may appear computationally taxing, it should be noted that the derivation of the model is required only once.  Ideally, the training process is performed to derive a computationally inexpensive model that can be widely utilized for a range of applications.

\begin{figure}
	\begin{center}
		\includegraphics[width=1.0\textwidth]{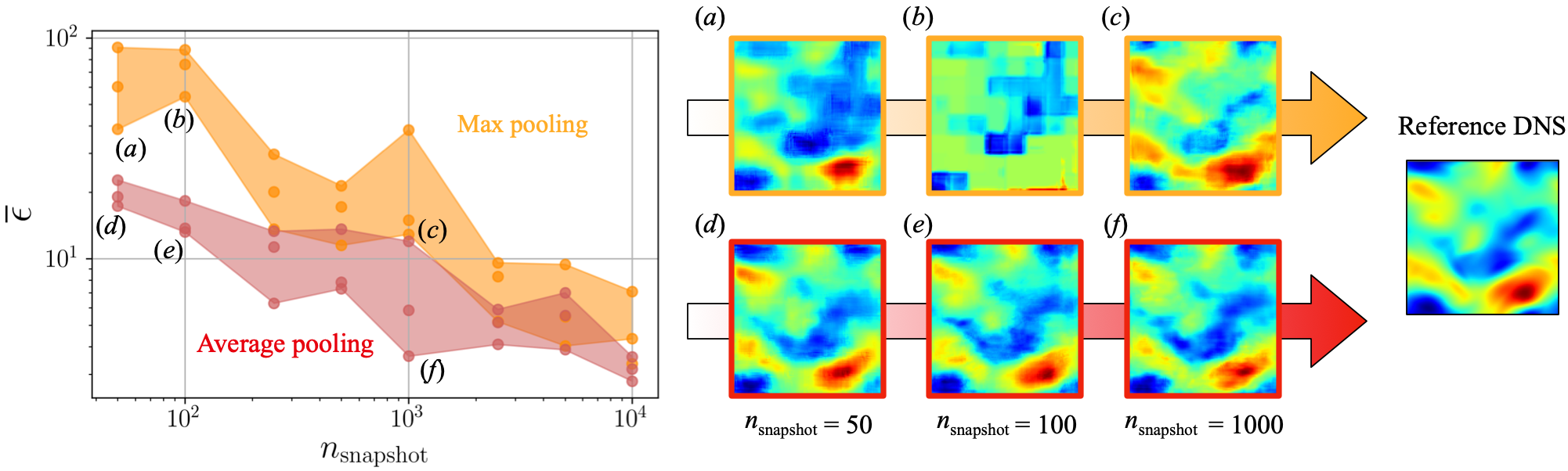}
		\caption{The dependence of the error $\overline{\epsilon}$ on snapshots $n_{\rm snapshot}$ of two-dimensional turbulence data. Shown on the right are the reconstructed velocity flow from low-resolution input data sets.}
		\label{fig6}
	\end{center}
\end{figure}

To determine the appropriate weights $\bm w$ by learning from data sets, there should be a sufficient number of snapshots provided to the CNN and hybrid DSC/MS models.  The influence of the number of snapshots on the accuracy of the reconstruction is assessed by evaluating the error for $n_{\rm snapshot} = 50-10\,000$. Also, these low-resolution snapshots are selected at even time intervals from the training data sets.
In general, the error norms decrease with increasing $n_{\rm snapshot}$ for the max and average pooled input data with the average pooled data consistently achieving lower level of errors, as shown in figure \ref{fig6}.  
Here, we also display the reconstructed flows for $n_{\rm snapshot} = 50$, $100$ and $1000$.  
It should be noted that even with a mere number of $50$ training snapshots, we can produce a reliable DSC/MS model to reconstruct the velocity field.

The distributions and characteristics of the input data are important in machine learning \citep{SHH1996}.  This can be said with regard to the accuracy achieved by the use of the max and average pooled data. For the max pooled data at $n_{\rm snapshot} = 100$, we find that the returned flow shows the staircased distribution, suggesting that the learning process is not stable and does not achieve a desirable trend in terms of convergence around $n_{\rm snapshot} = 100$.  Another aspect of the input data that needs some care is data standardization or normalization.  Within the context of our work, we utilized raw numerical values to perform learning but the values can be normalized to standardize the learning process.

The ability to reconstruct the subgrid-scale structures from the coarse data has many implications beyond simple flow reconstruction.  Not only do we desire to replicate the accurate statistics of turbulence but we also wish that the present approach provides possibilities for guiding future subgrid-scale models in turbulent flow simulations.  While we do not aim to reproduce high-order turbulent statistics in this study, we consider the accuracy of the current machine-learned model for the kinetic energy spectra.

\begin{figure}
	\begin{center}
		\includegraphics[width=0.9\textwidth]{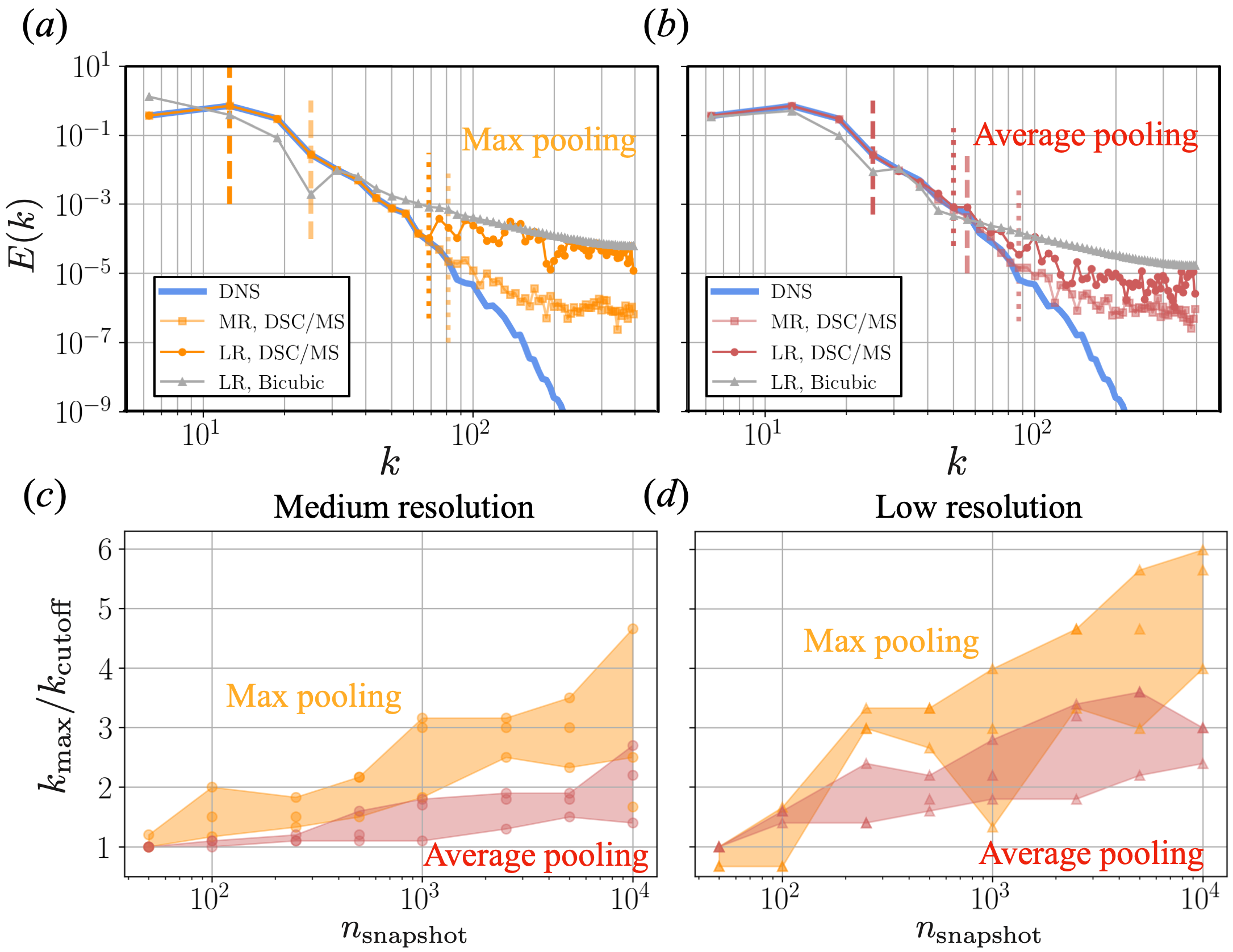}
		\caption{Kinetic energy spectra for $(a)$ max and $(b)$ average pooled data. Dashed and dotted lines indicate $k_{\rm cutoff}$ and $k_{\rm max}$, respectively, for medium and low-resolution data sets., The influence of $n_{\rm snapshot}$ on the ratio $k_{\rm max}/k_{\rm cutoff}$ for $(c)$ medium and $(d)$ low-resolution data, respectively.}
		\label{fig7}
	\end{center}
\end{figure}

Let us present the kinetic energy spectra from the hybrid DSC/MS model using the medium (MR) and low-resolution (LR) max and average pooled data in figure \ref{fig7}($a$).  
Here, we observe that the hybrid DSC/MS model reproduces the kinetic energy spectra over the spatial wavenumbers $k$ in an accurate manner.  
Dashed and dotted vertical lines in figure \ref{fig7}($a$) indicate the cutoff wavenumber $k_{\rm cutoff}$ for the MR and LR data, and the maximum wavenumber $k_{\rm max}$ in the reconstructed data, up to where the kinetic energy profile from the super-resolution solution matches at least $90\%$ with the reference profile.  
For comparison, we also present the results from the bicubic interpolation, which shows accuracy degradation for low-resolution input data, as also seen in figure \ref{fig4}. 
While the use of the max pooled data can recover the kinetic energy spectra beyond the bicubic interpolation, we see further improvement with the use of the average pooled input data for all cases.  We note that the machine-learned model cannot predict the energy spectra below $E(k) \approx 10^{-5}$. This is likely caused by the loss of data correlation between the large (input) and small-scale structures over a gap in wavenumbers.

Compiled in figure \ref{fig7}($b$) are the ratios $k_{\rm max}/k_{\rm cutoff}$ from reconstructed flow fields from the hybrid DSC/MS model to that of the medium and low-resolution input data.  With increasing snapshots of training data, we find that the kinetic energy spectra is recovered well.  For a given resolution of the input data, this figure can uncover the necessary number of snapshots to achieve the desired level of super-resolution enhancement.  What is noteworthy here is that with max pooling the recovery ratio can be over five fold in some cases.  For the average pooled input data, we can achieve over two fold in increasing $k_{\rm max}$.  
While max pooling does offer a high-level of super-resolution reconstruction, we should point out that the original error level is high to begin with.

\begin{figure}
	\begin{center}
		\includegraphics[width=1.0\textwidth]{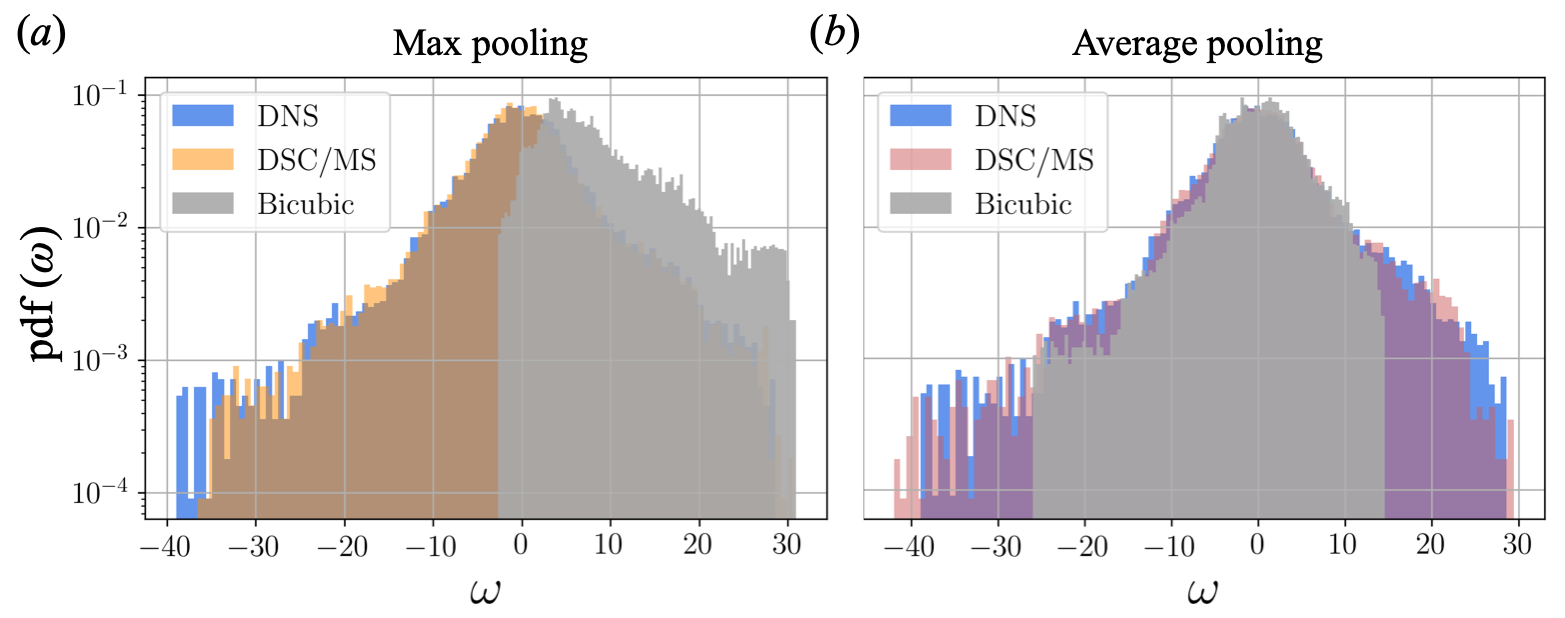}
		\caption{Probability density function of the vorticity field pdf$(\omega)$ for reference DNS, bicubic interpolation, and the hybrid DSC/MS model with low-resolution $(a)$ max and $(b)$ average pooling.}
		\label{fig9}
	\end{center}
\end{figure}

Moreover, we statistically examine the probability density function of the reconstructed vorticity field $\text{pdf}(\omega)$, as shown in figure \ref{fig9}. The low-resolution max and average pooled coarse vorticity fields are used as the input data. With max pooling, the bicubic interpolation is not able to recover the correct range of the vorticity field. On the other hand, the vorticity field is recovered well by using the hybrid DSC/MS model as shown by the orange area of figure \ref{fig9}($a$). With the average pooled data, the bicubic interpolation shows modest improvement over the max pooled case as shown in figure \ref{fig9}($b$).  The hybrid DSC/MS model can however accurately recover the flow field in a statistical manner for both input types, capturing the correct ranges of the vorticity field.

These results suggest the potential for accurately reconstructing and estimating important turbulent flow quantities, including the eddy viscosity coefficient. 
As seen from these results, the super-resolution analysis can provide insights that may have been considered unattainable from traditional approaches.  With the gargantuan amount of turbulent flow data from numerical simulations and experiments being stored and available, we should be able to utilize such library of data to perform super-resolution analysis of coarse flow field data for a range of complex turbulent flows.

\section{Conclusion}
We considered two machine-learning based approaches to perform super-resolution reconstruction of coarse flow fields.  The standard CNN was first studied and an improved hybrid DSC/MS model that handles the multi-scale nature of the flow was developed.  Both models were able to reconstruct laminar and turbulent flows.  
The two-dimensional cylinder wake was considered as the first example to show the overall super-resolution process and demonstrate its ability on laminar flow data.
Moreover, the two super-resolution models were assessed in detail for a canonical problem of two-dimensional decaying homogeneous turbulence.  The kinetic energy spectra can also be accurately reproduced.  The hybrid DSC/MS model was found to accurately reconstruct turbulent velocity and vorticity fields from extremely low-resolution input data. We believe that these models will perform even better for statistically stationary turbulent flows.   The dependence of the number of snapshots on the accuracy of the reconstruction was analyzed.  In some cases, the model can be learned with as little as 50 snapshots of training data. The influence of different coarsening techniques was also examined.  The average pooling method was found to be more robust than the max pooling method, which is prevalent in image processing.  With the ability to reconstruct the subgrid flow field with these machine-learned models, we should be able to extract physical insights beyond those directly estimated from the coarse data.  
 The purpose of the exercise presented in the paper was to consider the coarse flow field as images to reconstruct the subgrid scale structures.  During the machine-learned super-resolution (reconstruction) process, we did not assume a priori knowledge of the governing equations or statistical properties.  We believe that this is very important as a first step in demonstrating the strength of machine-learned techniques for super-resolution analysis.  This raises an important question of whether we can incorporate the governing equations into the learning process in future studies.
 With the increasing volume of high-fidelity reference data for a variety of flows, we should be able to utilize those big data to construct reliable machine-learned models to perform super-resolution analysis of a range of flows.

\section*{Acknowledgements}
K. Fukami and K. Fukagata thank the support from Japan Society for the Promotion of Science (KAKENHI grant number: 18H03758). K. Taira acknowledges the support from the US Army Research Office (grant number: W911NF-17-1-0118) and US Air Force Office of Scientific Research (grant number: FA9550-16-1-0650) and thanks L.~Mathelin, S.~L.~Brunton and J.~N.~Kutz for the stimulating discussions.

\bibliographystyle{jfm}

\end{document}